\documentclass{article}
\usepackage{amsmath,graphicx,mlspconf}
\usepackage{booktabs,array,url}
\usepackage{xcolor}
\usepackage{hyperref}

\title{Transfer Learning for Avian Bioacoustics under Sparse Positive Labels}

\name{Dhyey Patel, Yunting Yin}
\address{%
    Eastern Michigan University \\
    Computer Science Department \\
    Ypsilanti, MI, USA \\
    \texttt{dpatel48@emich.edu, yyin@emich.edu}%
}

\begin{document}
\ninept
\sloppy
\maketitle

\begin{abstract}
Passive acoustic monitoring is an important tool for biodiversity assessment and wildlife conservation because it supports continuous and non-invasive monitoring of species across large spatial and temporal scales. Robust monitoring remains challenging because many datasets contain sparse positive labels, where species presences may be confirmed while unannotated species cannot be assumed absent. In this work, we study transfer learning under sparse positive labels using BirdCLEF+ 2026 as a target benchmark and BirdCLEF 2021, iNatSounds, WABAD, and BirdSet as external bioacoustic sources. We introduce a multi-source reliability framework that models heterogeneous bioacoustic datasets as distinct supervision sources with differing reliability. Our approach achieves 0.584 macro average precision and 0.860 macro AUC on public BirdCLEF+ 2026 validation labels while outperforming naive source pooling strategies. The strongest gains arise from passive acoustic monitoring datasets and biologically informed source selection. Our findings suggest that transfer learning in bioacoustics is fundamentally a weak supervision and negative transfer problem. Source code is available at \href{https://github.com/AcaiLab/BirdMLSP}{https://github.com/AcaiLab/BirdMLSP}.
\end{abstract}


\section{Introduction}
Passive acoustic monitoring is a powerful approach for studying biodiversity and supporting wildlife conservation \cite{teixeira2024effective}. By continuously recording environmental soundscapes, researchers can monitor species presence, distribution, and activity patterns across large geographic regions and long time periods without disturbing natural habitats. The growing deployment of autonomous recording devices has produced vast collections of bioacoustic data, creating both opportunities and challenges for automated species recognition. Recent advances in machine learning and deep learning have significantly improved the ability to identify species from environmental recordings.

Despite this progress, developing robust bioacoustic recognition systems remains challenging. A fundamental difficulty is that many passive acoustic monitoring datasets contain incomplete annotations. In practice, a recording may be labeled with species that were confirmed to be present, while other vocalizing species remain unlabeled. As a result, the absence of a label cannot be interpreted as confirmed species absence. This creates a sparse positive label setting in which recorded presences are known but many negatives are uncertain. Such data violate assumptions commonly made in supervised learning and transfer learning, where unlabeled examples are often treated as negative examples.

Transfer learning has become an important strategy in bioacoustics since large external datasets are increasingly available. These datasets contain thousands of hours of recordings spanning diverse species and habitats. However, these datasets differ substantially in geographic coverage and taxonomic composition. Transferring knowledge from external datasets is not simply a matter of adding more training data. In this work, we study transfer learning under sparse positive labels using BirdCLEF+ 2026 \cite{birdclef2026} as a target benchmark and four external bioacoustic sources: BirdCLEF 2021 \cite{birdclef2021}, iNatSounds \cite{chasmai2024inatsounds}, WABAD \cite{wabad}, and BirdSet \cite{rauch2024birdset}. We formulate multi-source transfer as a positive-unlabeled learning problem in which external datasets contribute weak positive supervision.

Our study is motivated by two research questions. First, how should external bioacoustic datasets be incorporated when labels are sparse and incomplete? Second, when does transfer learning improve recognition performance, and when does it lead to negative transfer? To answer these questions, we systematically evaluate the effects of ecological information, feature representation learning, and transfer learning strategies on recognition performance under sparse positive labels. We find that negative transfer is common in sparse-label bioacoustic recognition. External datasets frequently improve performance on the species they cover while simultaneously reducing performance across the full target label set. We also find that transfer effectiveness depends strongly on ecological similarity and label coverage. Furthermore, treating heterogeneous datasets as a single source of supervision is often less effective than modeling their differences explicitly. To address these challenges, we introduce a multi-source reliability framework that treats external datasets as distinct sources of supervision with varying levels of reliability and transfer utility. Our proposed framework consistently reduces the impact of negative transfer and provides a reliable mechanism for integrating heterogeneous bioacoustic data.

\section{Related Work}
Bird audio recognition has benefited from the development of large-scale datasets and deep learning models. Benchmark resources such as BirdNET \cite{birdnet}, BirdSet \cite{rauch2024birdset}, WABAD \cite{wabad}, and the Benchmark of Animal Sounds \cite{beans} have enabled the training and evaluation of better recognition systems. Pretrained audio models including the Audio Spectrogram Transformer (AST) \cite{gong_ast}, Whisper \cite{whisper}, and Wav2Vec2 \cite{wav2vec2} have further improved acoustic representation learning across a variety of audio classification tasks. Researchers found that passive acoustic monitoring paired with machine learning can outperform traditional survey methods for detecting rare and cryptic species \cite{kurtin2025cuckoo}. Recent work on domain-invariant bird sound embeddings and global birdsong representations has demonstrated that external bioacoustic datasets can provide useful transfer signal \cite{moummad2024domain,ghani2023global}.

Transfer learning is particularly useful in bioacoustics because collecting labeled biodiversity audio is expensive. Day and Khoshgoftaar \cite{day2017heterogeneous} surveyed transfer learning across heterogeneous domains and highlighted challenges arising from differences in feature spaces, label spaces, and data distributions. More broadly, transfer learning literature has identified negative transfer as a major concern when source domains are insufficiently related to the target task \cite{cui2020multisource}. Weak and multilabel supervision are central to passive acoustic monitoring. Species may vocalize only briefly within a recording, multiple species may overlap, and annotations are often incomplete \cite{briggs2012acoustic,troshani2024weak}. Our work connects these three research directions and treats heterogeneous bioacoustic datasets as distinct sources of weak positive supervision.

\section{Datasets}
\begin{figure}[t]
\centering
\includegraphics[width=\linewidth]{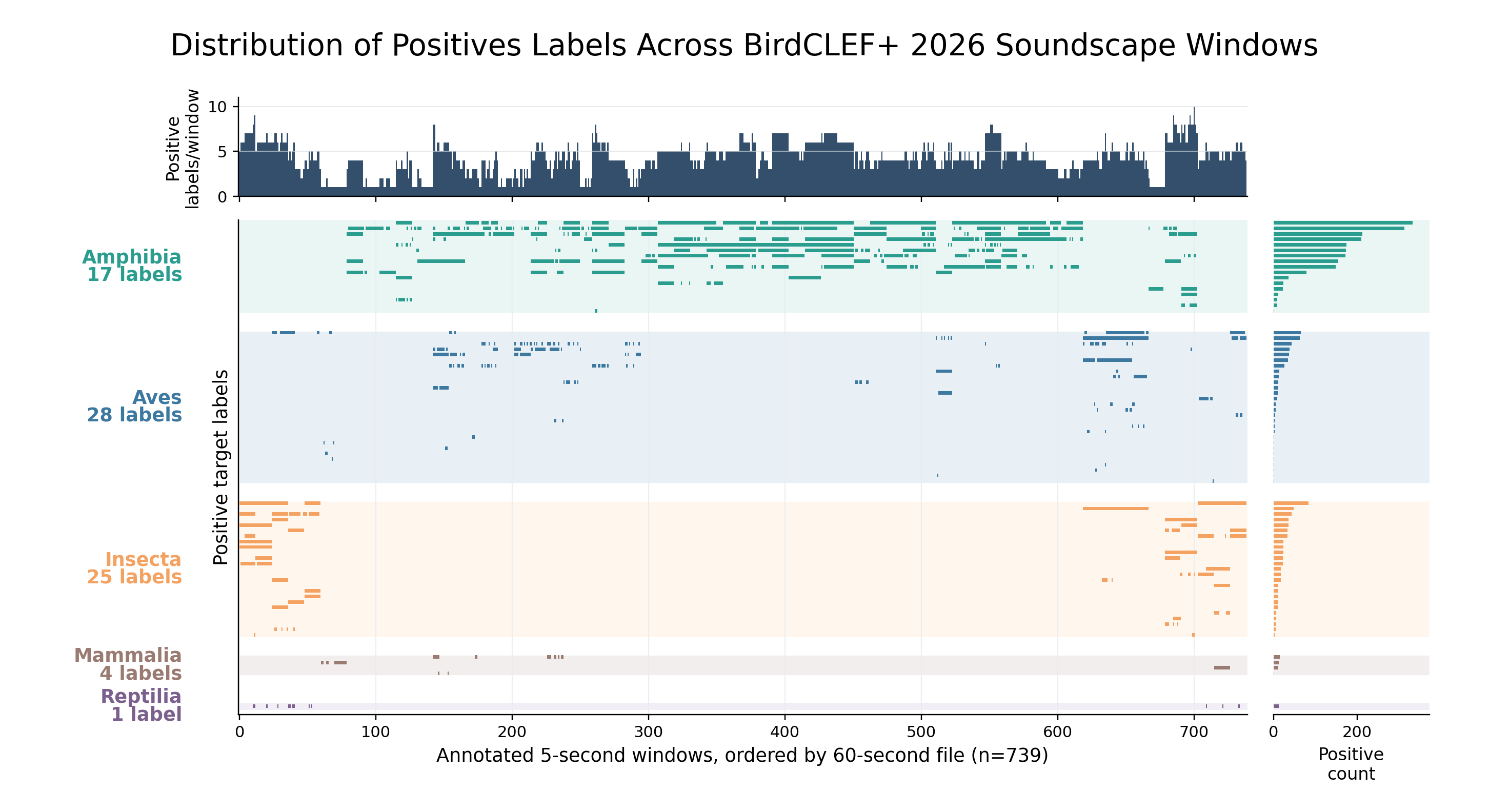}
\caption{Distribution of public positive labels in BirdCLEF+ 2026 soundscape windows.}
\label{fig:sparse-positive-matrix}
\end{figure}

Table~\ref{tab:data} summarizes the datasets used in this study. BirdCLEF+ 2026 \cite{birdclef2026} serves as the target benchmark and contains 35,549 focal recordings, 10,658 60-second soundscapes, and 234 target labels. Each public soundscape is divided into twelve non-overlapping 5-second windows. The public soundscape-label table contains 1,478 raw rows, which reduce to 739 unique annotated windows after exact duplicate removal.
A key characteristic of BirdCLEF+ 2026 is the extreme sparsity of its public annotations. As shown in Figure~\ref{fig:sparse-positive-matrix}, the public validation surface contains only 3,122 positive window-label pairs across 739 annotated windows, and only 75 of the 234 target labels have at least one public positive example. The remaining 159 labels have no public positives, resulting in an overall label density of only 1.81\%. Validation folds are assigned at the 60-second file level so that adjacent windows from the same recording never appear on both sides of a split.

\begin{table}[t]
\centering
\caption{Datasets used for target validation and transfer. BirdCLEF+ 2026 is
the only evaluation target; all external data are mapped by normalized
scientific name and used as positive-only transfer evidence.}
\label{tab:data}
\resizebox{\linewidth}{!}{%
\begin{tabular}{lrrl}
\toprule
Source & Overlap labels & Samples & Role \\
\midrule
BirdCLEF+ 2026 \cite{birdclef2026} & 234 & 739 windows & Target \\
BirdCLEF 2021 \cite{birdclef2021} & 34 & 7,380 focal recordings & Transfer \\
iNatSounds \cite{chasmai2024inatsounds} & 183 & 4,002 train recordings & Transfer \\
WABAD \cite{wabad} & 96 & 8,887 windows & Transfer \\
BirdSet PER \cite{rauch2024birdset} & 23 & 7,705 windows & Transfer \\
BirdSet NES \cite{rauch2024birdset} & 19 & 1,223 positives & Transfer \\
\bottomrule
\end{tabular}
}
\end{table}

We study transfer from four external sources: BirdCLEF 2021, with focal recordings and ecological metadata \cite{birdclef2021}; iNatSounds, a broad community upload collection \cite{chasmai2024inatsounds}; WABAD, a passive acoustic monitoring dataset \cite{wabad}; and the PER and NES Neotropical soundscapes from BirdSet \cite{rauch2024birdset}. Their respective overlaps with the target are 34, 183, 96, 23, and 19 labels. Because iNatSounds lacks a uniform expert verification, we estimate its reliability using exact scientific name matching, cap-per-label sampling, training weights based on sources, and covered-label evaluation.

We report five-fold means for macro average precision, macro area under the receiver operating characteristic curve, micro average precision, and micro F1 score when a decision threshold is relevant. Macro AP is the primary ranking metric because it weights each species equally despite severe class imbalance.

\section{Methods}

\subsection{BirdCLEF-Only Models}

We first establish a set of BirdCLEF-only baselines that use no external transfer data. These models characterize the predictive value of ecological context, acoustic features, and various pretrained audio representations using only the target benchmark.

To quantify the contribution of ecological information, we construct smoothed priors based on site, month, hour, and their combinations. For a metadata group $g$, the smoothed ecological prior for label $\ell$ is

\[
\widehat{p}_{\ell}(g)=
\frac{n_{\ell,g}+\alpha \widehat{p}_{\ell}}
{n_g+\alpha},
\]

where $n_{\ell,g}$ is the number of positive training windows for label $\ell$ in group $g$, $n_g$ is the number of training windows in that group, $\widehat{p}_{\ell}$ is the global training prevalence of label $\ell$, and $\alpha$ determines how strongly the group estimate is pulled toward this overall prevalence.

Acoustic baselines include logistic regression and random forests trained on lightweight acoustic descriptors, including root mean square energy, zero-crossing rate, spectral centroid, bandwidth, rolloff, flatness, and band-energy ratios. We additionally evaluate mel-spectrogram features and PCA-compressed acoustic representations.

To assess the value of pretrained representations, we extract frozen embeddings from AST/AudioSet \cite{gong_ast}, Whisper-tiny \cite{whisper}, and Wav2Vec2-base \cite{wav2vec2} models and train one-versus-rest logistic classifiers on the resulting feature vectors.

\subsection{Positive-Unlabeled Transfer Learning}
The central challenge addressed in this work is that passive acoustic monitoring datasets often contain sparse positive labels. A recorded species presence is known, but omitted species cannot be assumed absent. We therefore use external datasets as weak sources of positive supervision instead of treating their labels as complete.

When an external recording is matched to a BirdCLEF target species through scientific name normalization, it is added as a positive example for that species only. Remaining target labels are treated as unknown. This distinction is particularly important for focal recordings and community-upload datasets such as iNatSounds, where the annotated species may be known while additional vocalizing species remain unlabeled.

To account for annotation uncertainty, we evaluate positive-unlabeled (PU) weighting schemes that reduce the penalty assigned to unlabeled species-label pairs. The PU objective is

\[
\mathcal{L}_{\mathrm{PU}}
=
-\sum_{i,\ell}
\left[
y_{i\ell}\log s_{i\ell}
+
\lambda(1-y_{i\ell})\log(1-s_{i\ell})
\right],
\]

where $y_{i\ell}$ denotes the public positive indicator, $s_{i\ell}$ is the predicted score, and $0<\lambda<1$ reduces the contribution of unlabeled examples relative to confirmed positives.

\subsection{Multi-Source Reliability Framework}

A primary goal of this study is to understand how heterogeneous external datasets should be incorporated under sparse positive labels. To this end, we evaluate transfer from BirdCLEF 2021, iNatSounds, WABAD, BirdSet PER, and BirdSet NES using source-specific transfer protocols and weighting strategies.

Beyond direct transfer, we introduce a multi-source reliability framework that learns how to combine predictions from target-only and source-specific models while retaining the identity of each prediction stream. All inputs are out-of-fold predictions, so each training window is scored by base models that were not trained on that window. Let \(\mathcal K\) denote all prediction streams used in a model variant and \(\mathcal E\subseteq\mathcal K\) its external streams. For window \(i\), species \(\ell\), and stream \(k\), \(z_{i\ell}^{(k)}\) is the predicted probability, clipped to \([10^{-6},1-10^{-6}]\) before conversion to log-odds. The feature vector groups the stream predictions, source-comparison features, recording context, and species information:
\[
\begin{aligned}
x_{i\ell}=\Big[&
\{z_{i\ell}^{(k)},\operatorname{logit}(z_{i\ell}^{(k)})\}_{k\in\mathcal K},\\
&\{z_{i\ell}^{(k)}-z_{i\ell}^{(\mathrm{AST})},
z_{i\ell}^{(k)}-z_{i\ell}^{(\mathrm{stack})},
c_\ell^{(k)},c_\ell^{(k)}z_{i\ell}^{(k)},\\
&\qquad c_\ell^{(k)}\operatorname{logit}(z_{i\ell}^{(k)})\}_{k\in\mathcal E},\\
&d_i,\ u_\ell
\Big]^\top .
\end{aligned}
\]
Here, \(c_\ell^{(k)}=1\) when source \(k\) contains species \(\ell\). The context vector \(d_i\) contains scaled month, hour, and window start time. The species vector \(u_\ell\) contains training-fold prevalence and log positive count, source-coverage indicators, the number of covering sources, and one-hot taxonomic class. The resulting dimension is \(2|\mathcal K|+5|\mathcal E|+18\): 24 features for the target-only model, 31 with one external source, 52 for legacy or leave-one-source variants, and 59 with all sources.

Within each fold, we fit a standardizer using only the training pairs and apply it to both training and held-out pairs, producing \(\widetilde{x}_{i\ell}\). An L2-regularized logistic meta-classifier then produces the final probability:
\[
\begin{aligned}
q_{i\ell}&=\sigma(\beta_0+\beta^\top\widetilde{x}_{i\ell}),\\
\mathcal J(\beta_0,\beta)&={}
-\sum_{i,\ell}a_{i\ell}\big[y_{i\ell}\log q_{i\ell}\\
&\quad +(1-y_{i\ell})\log(1-q_{i\ell})\big]
+\gamma\lVert\beta\rVert_2^2,\\
(\widehat{\beta}_0,\widehat{\beta})&=
\underset{\beta_0,\beta}{\operatorname{arg\,min}}\,
\mathcal J(\beta_0,\beta).
\end{aligned}
\]
Here, \(y_{i\ell}=1\) is a confirmed positive, while \(y_{i\ell}=0\) is an unobserved label rather than a confirmed absence. In PU variants, confirmed positives receive weight 1 and unobserved pairs receive weight \(\rho\in\{0.05,0.2\}\). Variants without PU weighting use scikit-learn's balanced class weights. We implement the L2 penalty with \(C=0.5\) and optimize with the \texttt{lbfgs} solver.

To capture label dependencies, we construct positive pointwise mutual information (PPMI) co-occurrence graphs using training-fold statistics only. For labels $\ell$ and $m$, the graph edge weight is

\[
\operatorname{PPMI}(\ell,m)=
\max\left(
0,
\log
\frac{\widehat{P}(\ell,m)}
{\widehat{P}(\ell)\widehat{P}(m)}
\right),
\]

where probabilities are estimated from training-fold co-occurrence counts. Positive values indicate label pairs that occur together more frequently than expected from their marginal frequencies.

The resulting framework allows the model to learn reliability patterns unique to each source and to distinguish between sources that provide useful transfer signal and those that introduce negative transfer. We also evaluate methods that incorporate temporal context across the twelve consecutive 5-second windows within each soundscape, including sequence-based calibration and a Markov-logit temporal smoother.

\subsection{Evaluation Protocol}

Performance is evaluated using macro average precision, macro area under the receiver operating characteristic curve, micro average precision, and thresholded micro F1. Because many transfer sources overlap with only a subset of BirdCLEF labels, results are reported for both all label evaluation and source-covered subsets.

To characterize transfer behavior, we additionally evaluate performance across label-frequency strata, taxonomic subgroups, and alternative calibration metrics including Brier score, log loss, and expected calibration error. Statistical uncertainty is quantified using file-level bootstrap confidence intervals and paired file-level permutation tests.

\section{Results}

\subsection{BirdCLEF-Only Baselines}

We begin by evaluating BirdCLEF-only baselines to establish the effects of ecological context, acoustic representations, and label co-occurrence. Table~\ref{tab:birdclef} summarizes the BirdCLEF-only results. Ecological context alone provides a surprisingly strong signal. While a global prevalence prior achieves only 0.127 macro AP, the site/month/hour prior reaches 0.462 macro AP, indicating that species occurrence patterns are strongly structured by ecological context. Control experiments that permute metadata assignments or break label-context relationships substantially reduce performance, suggesting that the gain reflects meaningful ecological information.

\begin{table}[t]
\centering
\caption{BirdCLEF+ 2026 public-label baselines and compact component grid.
AST denotes Audio Spectrogram Transformer embeddings, PPMI denotes positive
pointwise mutual information graph features, and PU denotes positive-unlabeled
negative weighting.}
\label{tab:birdclef}
\resizebox{\linewidth}{!}{%
\begin{tabular}{lrrrr}
\toprule
Model & Macro AP & Macro AUC & Micro AP & Micro F1 \\
\midrule
Global prior & 0.127 & 0.500 & 0.292 & 0.000 \\
Site/month/hour prior & 0.462 & 0.782 & 0.481 & 0.389 \\
Acoustic logistic regression & 0.440 & 0.783 & 0.389 & 0.455 \\
Acoustic random forest & 0.474 & 0.791 & 0.559 & 0.550 \\
Stacked audio-context & 0.527 & 0.822 & 0.598 & 0.376 \\
PPMI co-occurrence calibrator & 0.522 & 0.832 & 0.564 & 0.313 \\
PU-weighted PPMI calibrator & 0.521 & 0.832 & 0.586 & 0.534 \\
AST/AudioSet logistic & 0.491 & 0.770 & 0.630 & 0.596 \\
AST/AudioSet + PU co-occurrence & 0.515 & 0.825 & 0.590 & 0.590 \\
AST+stack fusion & 0.554 & 0.822 & 0.655 & 0.368 \\
AST+stack+PU fusion & 0.550 & 0.823 & 0.664 & 0.597 \\
\textbf{AST+stack+graph fusion} & \textbf{0.555} & \textbf{0.832} & \textbf{0.670} & \textbf{0.361} \\
AST+stack+graph+PU fusion & 0.549 & 0.833 & 0.673 & 0.596 \\
\bottomrule
\end{tabular}
}
\end{table}

\begin{figure}[t]
\centering
\includegraphics[width=\linewidth]{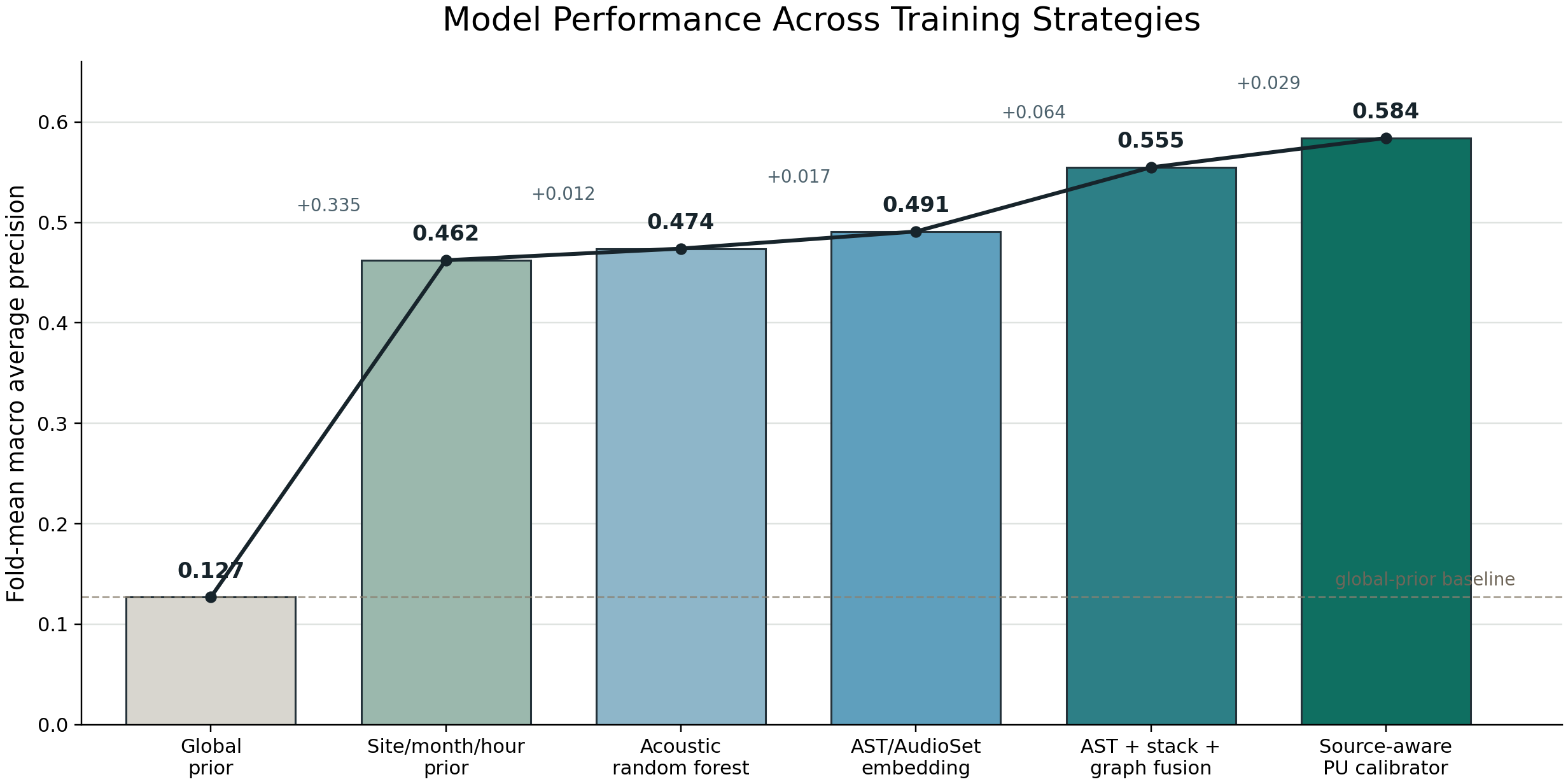}
\caption{Performance progression from ecological priors to multi-source reliability modeling on BirdCLEF+ 2026.}
\label{fig:baseline-progression}
\end{figure}

Acoustic modeling provides additional improvements. Random forests outperform linear acoustic baselines, and the stacked audio-context model reaches 0.527 macro AP. Among pretrained representations, AST embeddings provide the strongest standalone performance, achieving 0.491 macro AP and substantially outperforming Wav2Vec2-base. Combining AST predictions with stacked contextual features and label co-occurrence information further improves performance to 0.555 macro AP. Figure~\ref{fig:baseline-progression} summarizes this progression. The largest performance gain arises from ecological context, followed by smaller but consistent improvements from acoustic modeling, pretrained representations, and feature fusion. These results establish a strong target-only baseline before incorporating external transfer sources.

\begin{figure}[t]
\centering
\includegraphics[width=\linewidth]{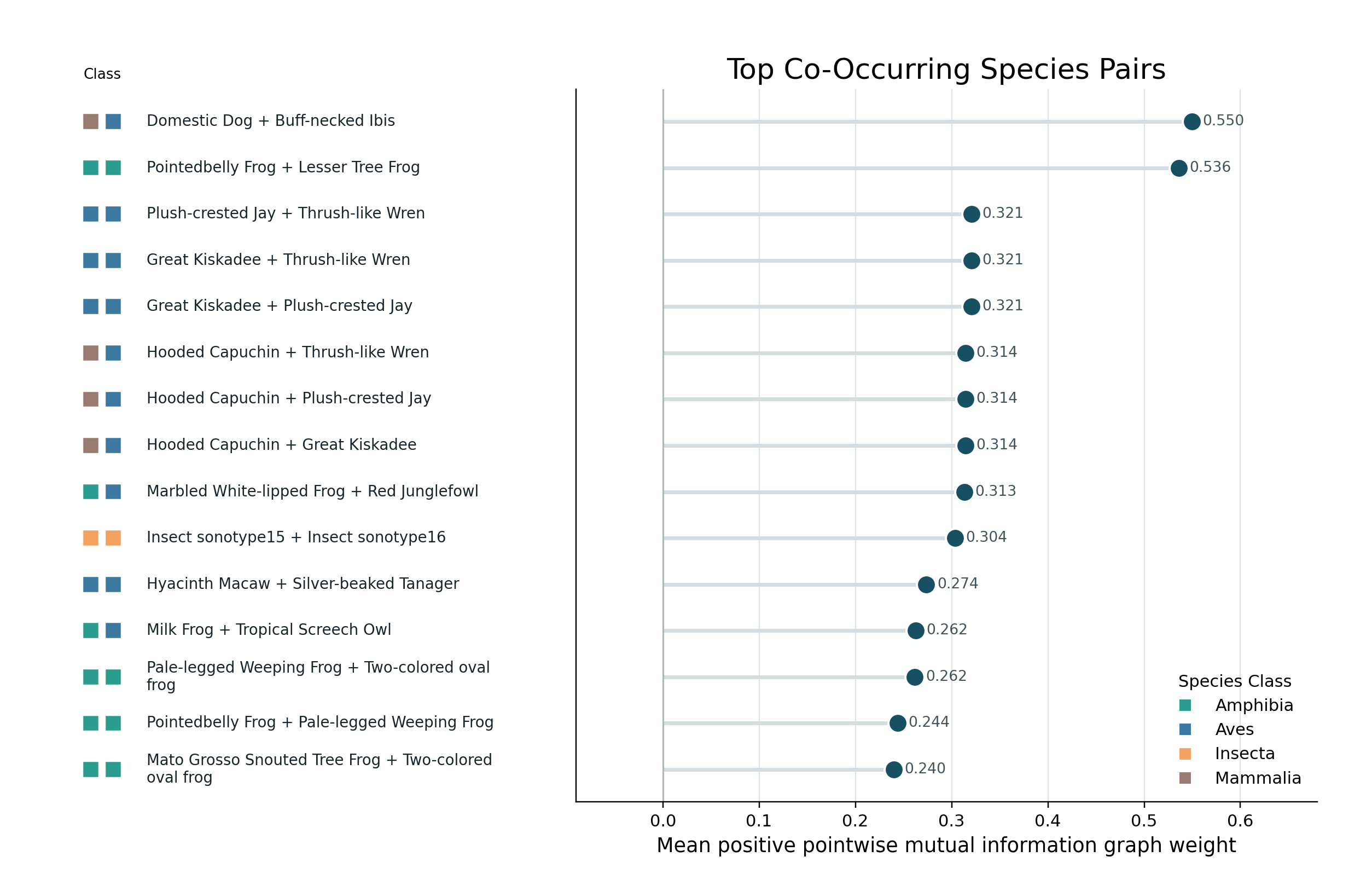}
\caption{Top co-occurring species pairs identified from training folds. Pairs are ranked by positive pointwise mutual information (PPMI), with longer bars indicating stronger co-occurrence. Colored markers denote the taxonomic classes of the two species.}
\label{fig:cooccurrence}
\end{figure}

Label co-occurrence provides useful but limited information. PPMI-based graph features consistently outperform shuffled controls, indicating that meaningful acoustic community structure exists within the data. However, graph features alone do not surpass the strongest audio-context models. 
Positive-unlabeled weighting improves micro F1 with little change in ranking performance, suggesting that co-occurrence information is most helpful for refining prediction scores, but not enough for driving the predictions themselves. Figure~\ref{fig:cooccurrence} illustrates the strongest learned co-occurrence relationships. These associations can be interpreted as shared acoustic or ecological context. Species that frequently co-occur often occupy similar habitats or recording conditions, making co-occurrence information useful even when it does not directly improve recognition accuracy.

\subsection{Transfer Learning Under Sparse Positive Labels}
We evaluate whether external datasets improve recognition performance under sparse positive labels. Table~\ref{tab:transfer} shows that external datasets often improve performance on source-covered labels without improving the full target label set. This pattern indicates negative transfer and motivates treating external datasets as distinct supervision sources.

\begin{figure}[t]
\centering
\includegraphics[width=\linewidth]{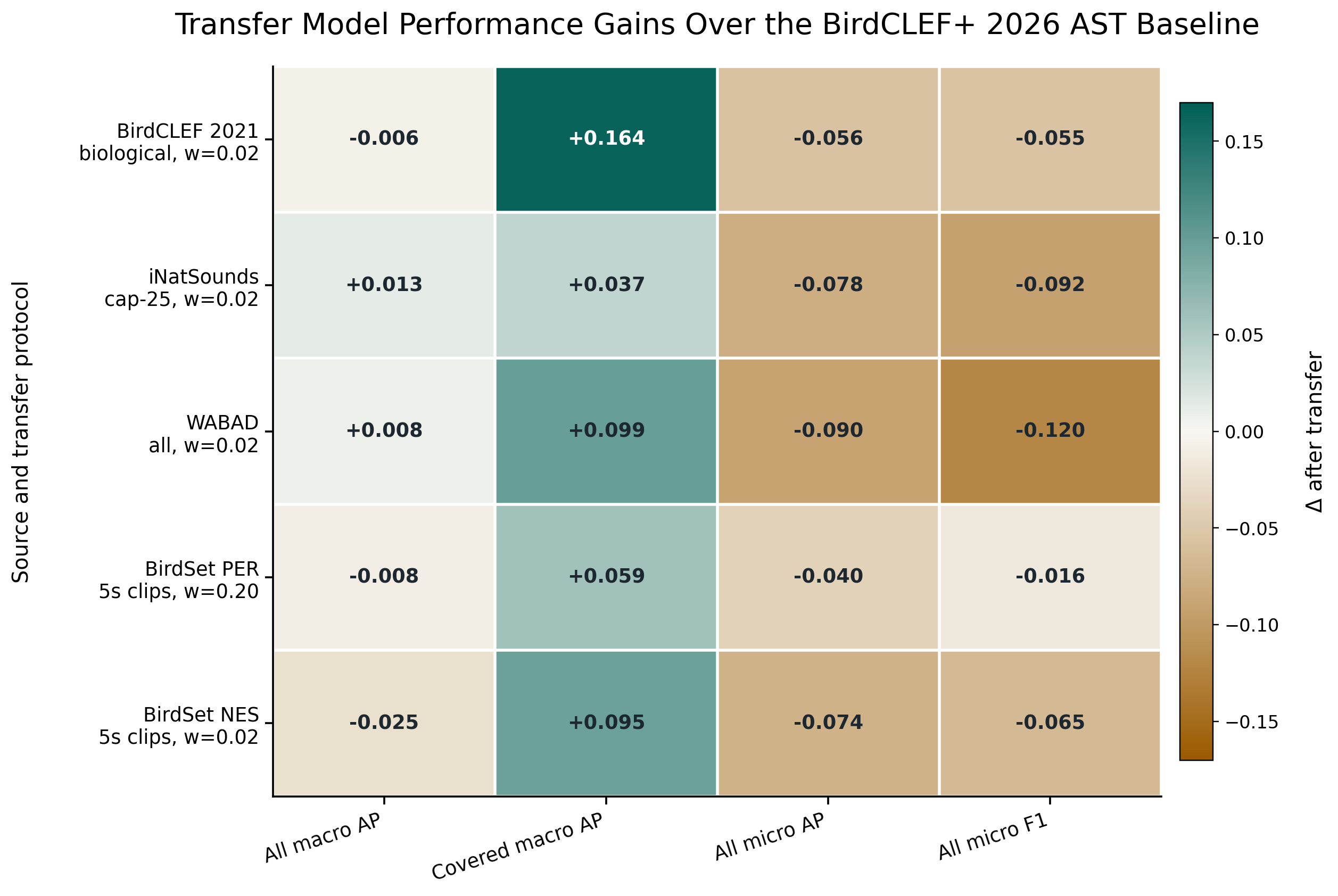}
\caption{Relative performance changes from external transfer. Overlap AP is evaluated on source-overlapping labels, whereas all-label metrics are computed on the full BirdCLEF+ 2026 benchmark.}
\label{fig:transfer-heatmap}
\end{figure}

Source relevance is critical here. For BirdCLEF 2021, unrestricted transfer reduces overlap performance, whereas filtering by season, geography, time, call type, and recording quality improves it. iNatSounds offers broader coverage but produces only modest covered-label gains. Passive acoustic monitoring sources transfer more effectively: WABAD increases overlap macro AP from 0.275 to 0.374, while BirdSet PER and NES improve from 0.176 to 0.235 and from 0.349 to 0.444, respectively. However, both BirdSet sources reduce all-label macro AP. Figure~\ref{fig:transfer-heatmap} illustrates this trade-off, showing why transfer must be evaluated on both covered labels and the full target set. This negative transfer likely reflects support mismatch. PER and NES cover only 23 and 19 target labels, with 48 and 133 public positive assignments, respectively. Their external positives can improve covered-label ranking while shifting predictions for the remaining 200+ labels. Focal recordings introduce an additional mismatch because their single annotated species does not establish the absence of other vocalizing species.

\begin{table}[t]
\centering
\caption{Transfer performance using external datasets as positive-only supervision sources. ``Overlap AP’’ is computed only on labels shared between BirdCLEF+ 2026 and the corresponding external dataset. Here \(w\) denotes the relative sample weight assigned to external positive examples compared with target BirdCLEF examples. ``Cap-\(k\)'' restricts each external label to at most \(k\) examples.}
\label{tab:transfer}
\resizebox{\linewidth}{!}{%
\begin{tabular}{llrrrr}
\toprule
Source & Setting & All AP & All micro AP & Overlap AP & Overlap micro AP \\
\midrule
2021 & 2026 acoustic only & 0.440 & 0.389 & 0.386 & -- \\
2021 & +50 focal/label & 0.443 & 0.421 & 0.422 & -- \\
2021 & 2026 AST only & 0.491 & 0.630 & 0.440 & 0.622 \\
2021 & \textbf{Biological AST, w=0.02} & \textbf{0.485} & \textbf{0.574} & \textbf{0.605} & \textbf{0.639} \\
\midrule
iNat & 2026 AST only & 0.491 & 0.630 & 0.480 & 0.699 \\
iNat & \textbf{iNat cap-25 AST, w=0.02} & \textbf{0.504} & \textbf{0.552} & \textbf{0.518} & \textbf{0.613} \\
\midrule
WABAD & 2026 AST only & 0.491 & 0.630 & 0.275 & 0.367 \\
WABAD & \textbf{WABAD all, w=0.02} & \textbf{0.499} & \textbf{0.540} & \textbf{0.374} & \textbf{0.323} \\
\midrule
PER & 2026 AST only & 0.491 & 0.630 & 0.176 & 0.170 \\
PER & \textbf{PER 5s all AST, w=0.20} & \textbf{0.483} & \textbf{0.590} & \textbf{0.235} & \textbf{0.154} \\
\midrule
NES & 2026 AST only & 0.491 & 0.630 & 0.349 & 0.472 \\
NES & \textbf{NES 5s all AST, w=0.02} & \textbf{0.466} & \textbf{0.556} & \textbf{0.444} & \textbf{0.367} \\
\bottomrule
\end{tabular}
}
\end{table}

\subsection{Mitigating Negative Transfer}
We investigate strategies for mitigating negative transfer. Table~\ref{tab:mitigation} shows that source selection matters more than the precise transfer weight, which has relatively little effect between 0.02 and 0.20 once an appropriate protocol is chosen. Relevance filtering is the most effective strategy: for example, biologically informed BirdCLEF 2021 sampling outperforms unrestricted transfer. However, gains on source-covered labels often coincide with poorer all-label performance, indicating distributional mismatch. Transfer effectiveness therefore depends more on ecological relevance, annotation completeness, and label coverage than on dataset size alone.

\begin{table}[t]
\centering
\caption{Standardized source-weight grid for negative-transfer mitigation.
Covered-label metrics are computed only on labels supported by the external
source. Within each source block, the best covered-label macro AP row is
bolded.}
\label{tab:mitigation}
\scriptsize
\resizebox{\linewidth}{!}{%
\begin{tabular}{llrrrr}
\toprule
Source & Protocol & All AP & All micro AP & Covered AP & Covered micro AP \\
\midrule
BirdCLEF 2021 & 2026-only AST & 0.491 & 0.630 & 0.440 & 0.622 \\
BirdCLEF 2021 & \textbf{Biological, w=0.02} & \textbf{0.485} & \textbf{0.574} & \textbf{0.605} & \textbf{0.639} \\
BirdCLEF 2021 & Biological, w=0.05 & 0.485 & 0.574 & 0.601 & 0.640 \\
BirdCLEF 2021 & Biological, w=0.10 & 0.485 & 0.574 & 0.602 & 0.642 \\
BirdCLEF 2021 & Biological, w=0.20 & 0.485 & 0.574 & 0.600 & 0.644 \\
\midrule
iNatSounds & 2026-only AST & 0.491 & 0.630 & 0.480 & 0.699 \\
iNatSounds & \textbf{Cap-25, w=0.02} & \textbf{0.504} & \textbf{0.552} & \textbf{0.518} & \textbf{0.613} \\
iNatSounds & Cap-25, w=0.05 & 0.504 & 0.553 & 0.517 & 0.613 \\
iNatSounds & Cap-25, w=0.10 & 0.504 & 0.553 & 0.517 & 0.613 \\
iNatSounds & Cap-25, w=0.20 & 0.504 & 0.553 & 0.517 & 0.614 \\
\midrule
WABAD & 2026-only AST & 0.491 & 0.630 & 0.275 & 0.367 \\
WABAD & \textbf{All windows, w=0.02} & \textbf{0.499} & \textbf{0.540} & \textbf{0.374} & \textbf{0.323} \\
WABAD & All windows, w=0.05 & 0.499 & 0.539 & 0.373 & 0.324 \\
WABAD & All windows, w=0.10 & 0.497 & 0.539 & 0.367 & 0.323 \\
WABAD & All windows, w=0.20 & 0.497 & 0.539 & 0.365 & 0.323 \\
\midrule
BirdSet PER & 2026-only AST & 0.491 & 0.630 & 0.176 & 0.170 \\
BirdSet PER & 5s clips, w=0.02 & 0.483 & 0.590 & 0.234 & 0.165 \\
BirdSet PER & 5s clips, w=0.05 & 0.483 & 0.590 & 0.233 & 0.158 \\
BirdSet PER & 5s clips, w=0.10 & 0.483 & 0.590 & 0.232 & 0.155 \\
BirdSet PER & \textbf{5s clips, w=0.20} & \textbf{0.483} & \textbf{0.590} & \textbf{0.235} & \textbf{0.154} \\
\midrule
BirdSet NES & 2026-only AST & 0.491 & 0.630 & 0.349 & 0.472 \\
BirdSet NES & \textbf{5s clips, w=0.02} & \textbf{0.466} & \textbf{0.556} & \textbf{0.444} & \textbf{0.367} \\
BirdSet NES & 5s clips, w=0.05 & 0.466 & 0.556 & 0.443 & 0.365 \\
BirdSet NES & 5s clips, w=0.10 & 0.466 & 0.556 & 0.443 & 0.364 \\
BirdSet NES & 5s clips, w=0.20 & 0.466 & 0.556 & 0.442 & 0.357 \\
\bottomrule
\end{tabular}
}
\end{table}

\section{Multi-Source Transfer Learning}

The previous experiments demonstrate that no single transfer source consistently improves all metrics. Different datasets help different label subsets, motivating a framework that explicitly models source reliability instead of pooling all external data into a single training set. Table~\ref{tab:final} summarizes the multi-source reliability experiments. Even without external predictions, the target-only reliability model improves performance over the strongest BirdCLEF-only baselines. This suggests that positive-unlabeled refinement and features that account for differences across data sources both improve the predictions. When individual external datasets are added, performance varies substantially across sources. WABAD and iNatSounds provide the strongest single-source improvements, whereas BirdSet PER and NES are less effective despite their overlap gains.

Another important finding is that combining all external datasets without considering their differences can reduce performance. The strongest fold-mean ranking model excludes BirdSet PER and achieves 0.584 macro AP. However, several nearby variants achieve nearly identical performance, including the legacy-source model and the leave-out-NES configuration. Pooling all sources is not optimal, as the all-source model reaches 0.578 macro AP, below several source-selected variants, despite improving some other metrics. These findings suggest that modeling source reliability is more important than simply increasing the number of transfer sources.

Temporal context provides a complementary signal. Sequence-based refinement improves micro F1 to 0.611, whereas explicit source modeling yields the strongest macro AP. This distinction suggests that temporal information is most useful for thresholded decision making, while source reliability primarily benefits class-balanced ranking.

We observe substantial label persistence across neighboring windows. The global adjacent-window Jaccard is 0.830 for BirdCLEF and 0.637 for matched WABAD passive windows. This motivates our use of low-capacity temporal score refinement. More expressive sequence models, such as CRFs, LSTMs, or attention-based models may provide further gains.

Not all observed performance gains are equally well supported by the uncertainty analysis. WABAD provides the clearest statistically supported transfer result, improving overlap macro AP by 0.069 (p=0.032). BirdCLEF 2021 biological sampling shows a similarly sized gain but does not reach conventional significance thresholds (p=0.064). PER and NES exhibit positive overlap gains, but uncertainty remains substantial due to limited overlap support. Likewise, the strongest reliability model variants are statistically indistinguishable from several nearby alternatives.  We therefore interpret the results as evidence that heterogeneous datasets should not be pooled indiscriminately.

\begin{table}[t]
\centering
\caption{Source-aware calibration ablation and uncertainty results. PU denotes
positive-unlabeled negative weighting. Single-source rows add one external
stream to the BirdCLEF-only prediction streams; leave-out rows include all
external streams except the named source. Best ranking row is bolded. The upper panel reports means across five folds; permutation-test deltas in the lower panel are computed from pooled out-of-fold predictions and therefore need not equal differences between the fold means.}
\label{tab:final}
\scriptsize
\resizebox{\linewidth}{!}{%
\begin{tabular}{lrrrr}
\toprule
Model & Macro AP & Macro AUC & Micro AP & Micro F1 \\
\midrule
AST+stack+graph fusion & 0.555 & 0.832 & 0.670 & 0.361 \\
AST+stack PU fusion & 0.550 & 0.823 & 0.664 & 0.597 \\
Markov-logit temporal smoother & 0.559 & 0.837 & 0.675 & 0.364 \\
Source-aware target-only, PU 0.2 & 0.572 & 0.826 & 0.629 & 0.608 \\
Source-aware + BirdCLEF 2021 only, PU 0.2 & 0.568 & 0.834 & 0.631 & 0.603 \\
Source-aware + iNatSounds only, PU 0.2 & 0.578 & 0.852 & 0.635 & 0.599 \\
Source-aware + WABAD only, PU 0.2 & 0.576 & 0.852 & 0.630 & 0.608 \\
Source-aware + BirdSet PER only, PU 0.2 & 0.557 & 0.829 & 0.628 & 0.603 \\
Source-aware + BirdSet NES only, PU 0.2 & 0.549 & 0.830 & 0.625 & 0.607 \\
Source-aware legacy sources, PU 0.2 & 0.583 & 0.858 & 0.639 & 0.602 \\
Source-aware all sources, PU 0.2 & 0.578 & 0.859 & 0.639 & 0.601 \\
Leave out BirdCLEF 2021, PU 0.2 & 0.580 & 0.859 & 0.636 & 0.597 \\
Leave out iNatSounds, PU 0.2 & 0.572 & 0.853 & 0.628 & 0.602 \\
Leave out WABAD, PU 0.2 & 0.573 & 0.850 & 0.634 & 0.601 \\
\textbf{Leave out BirdSet PER, PU 0.2} & \textbf{0.584} & \textbf{0.860} & \textbf{0.640} & \textbf{0.602} \\
Leave out BirdSet NES, PU 0.2 & 0.583 & 0.858 & 0.639 & 0.600 \\
Sequence on leave-out-PER source-aware model & 0.555 & 0.838 & 0.654 & 0.611 \\
\bottomrule
\end{tabular}
}
\vspace{0.3em}
\resizebox{\linewidth}{!}{%
\begin{tabular}{llrr}
\toprule
Paired comparison & Subset & $\Delta$ AP & Two-sided $p$ \\
\midrule
Leave-out-PER source-aware vs. AST+stack+graph PU & All & +0.007 & 0.526 \\
Sequence leave-out-PER vs. leave-out-PER source-aware & All & +0.015 & 0.048 \\
Sequence leave-out-PER vs. Markov smoother & All & +0.007 & 0.757 \\
BirdCLEF 2021 biological vs. 2026-only & 2021 overlap & +0.068 & 0.064 \\
iNatSounds cap-25 vs. 2026-only & iNatSounds overlap & +0.028 & 0.375 \\
\textbf{WABAD all vs. 2026-only} & \textbf{WABAD overlap} & \textbf{+0.069} & \textbf{0.032} \\
BirdSet PER 5s vs. 2026-only & PER overlap & +0.040 & 0.112 \\
BirdSet NES 5s vs. 2026-only & NES overlap & +0.161 & 0.502 \\
\bottomrule
\end{tabular}
}
\end{table}

\section{Conclusion}
We investigated transfer learning for BirdCLEF+ 2026 under sparse positive labels using ecological priors, pretrained audio representations, label co-occurrence modeling, positive-unlabeled learning, and heterogeneous external bioacoustic datasets. Ecological context is a strong predictor across a broad set of experiments. Pretrained audio representations and feature fusion provided additional improvements. The strongest overall performance was achieved by multi-source reliability modeling.

More importantly, our results suggest that transfer learning under sparse positive labels is fundamentally different from conventional supervised transfer learning. External datasets frequently improved performance on the species they covered, yet these gains did not consistently translate to improvements on the full target label set. Transfer effectiveness depended strongly on source characteristics, indicating that more external data is not necessarily better. Instead, effective transfer requires explicit consideration of annotation uncertainty, ecological similarity, and source reliability. A promising direction for future work is the development of soundscape-level models that jointly learn acoustic representations, temporal dependencies, ecological context, label co-occurrence, and source reliability within a unified learning framework.

\bibliographystyle{IEEEbib}
\bibliography{refs}

@misc{birdclef2026,
  author       = {{BirdCLEF}},
  title        = {{BirdCLEF+ 2026}},
  year         = {2026},
  howpublished = {Kaggle Competition Dataset},
  note         = {\url{https://www.kaggle.com/competitions/birdclef-2026}}
}

@misc{birdclef2021,
  author       = {{BirdCLEF}},
  title        = {{BirdCLEF 2021 Birdcall Identification}},
  year         = {2021},
  howpublished = {Kaggle Competition Dataset},
  note         = {\url{https://www.kaggle.com/c/birdclef-2021}}
}

@article{chasmai2024inatsounds,
      title={The iNaturalist Sounds Dataset},
      author={Chasmai, Mustafa and Shepard, Alex and Maji, Subhransu and Van Horn, Grant},
      journal={Advances in Neural Information Processing Systems},
      year={2024}
  }

@article{wabad,
author = {Pérez-Granados, Cristian and
          Morant, Jon and
          Darras, Kevin F. A. and
          Marín-Gómez, Oscar H. and
          Mendoza, Irene and
          Muñoz-Mohedano, Miguel A. and
          Santamaría-García, Eduardo and
          others},
title = {WABAD: A world annotated bird acoustic dataset for passive acoustic monitoring},
journal = {Ecology},
volume = {107},
number = {2},
pages = {e70317},
doi = {https://doi.org/10.1002/ecy.70317},
url = {https://esajournals.onlinelibrary.wiley.com/doi/abs/10.1002/ecy.70317},
eprint = {https://esajournals.onlinelibrary.wiley.com/doi/pdf/10.1002/ecy.70317},
year = {2026}
}

@misc{rauch2024birdset,
      title={BirdSet: A Large-Scale Dataset for Audio Classification in Avian Bioacoustics}, 
      author={Lukas Rauch and Raphael Schwinger and Moritz Wirth and René Heinrich and Denis Huseljic and Marek Herde and Jonas Lange and Stefan Kahl and Bernhard Sick and Sven Tomforde and Christoph Scholz},
      year={2024},
      eprint={2403.10380},
      archivePrefix={arXiv},
      primaryClass={cs.SD},
      url={https://arxiv.org/abs/2403.10380}, 
}

@article{birdnet,
title = {BirdNET: A deep learning solution for avian diversity monitoring},
journal = {Ecological Informatics},
volume = {61},
pages = {101236},
year = {2021},
issn = {1574-9541},
doi = {https://doi.org/10.1016/j.ecoinf.2021.101236},
url = {https://www.sciencedirect.com/science/article/pii/S1574954121000273},
author = {Stefan Kahl and Connor M. Wood and Maximilian Eibl and Holger Klinck},
}

@INPROCEEDINGS{beans,
  author={Hagiwara, Masato and Hoffman, Benjamin and Liu, Jen-Yu and Cusimano, Maddie and Effenberger, Felix and Zacarian, Katie},
  booktitle={ICASSP 2023 - 2023 IEEE International Conference on Acoustics, Speech and Signal Processing (ICASSP)}, 
  title={BEANS: The Benchmark of Animal Sounds}, 
  year={2023},
  volume={},
  number={},
  pages={1-5},
  doi={10.1109/ICASSP49357.2023.10096686}}

@inproceedings{gong_ast,
  title     = {{AST: Audio Spectrogram Transformer}},
  author    = {Yuan Gong and Yu-An Chung and James Glass},
  year      = {2021},
  booktitle = {{Interspeech 2021}},
  pages     = {571--575},
  doi       = {10.21437/Interspeech.2021-698},
  issn      = {2958-1796},
}

@INPROCEEDINGS{moummad2024domain,
  author={Moummad, Ilyass and Serizel, Romain and Benetos, Emmanouil and Farrugia, Nicolas},
  booktitle={ICASSP 2026 - 2026 IEEE International Conference on Acoustics, Speech and Signal Processing (ICASSP)}, 
  title={Domain-Invariant Representation Learning of Bird Sounds}, 
  year={2026},
  volume={},
  number={},
  pages={15237-15241},
  doi={10.1109/ICASSP55912.2026.11463533}}

@Article{ghani2023global,
author={Ghani, Burooj
and Denton, Tom
and Kahl, Stefan
and Klinck, Holger},
title={Global birdsong embeddings enable superior transfer learning for bioacoustic classification},
journal={Scientific Reports},
year={2023},
month={Dec},
day={18},
volume={13},
number={1},
pages={22876},
issn={2045-2322},
doi={10.1038/s41598-023-49989-z},
url={https://doi.org/10.1038/s41598-023-49989-z}
}

@article{briggs2012acoustic,
    author = {Briggs, Forrest and Lakshminarayanan, Balaji and Neal, Lawrence and Fern, Xiaoli Z. and Raich, Raviv and Hadley, Sarah J. K. and Hadley, Adam S. and Betts, Matthew G.},
    title = {Acoustic classification of multiple simultaneous bird species: A multi-instance multi-label approach},
    journal = {The Journal of the Acoustical Society of America},
    volume = {131},
    number = {6},
    pages = {4640-4650},
    year = {2012},
    month = {06},
    issn = {0001-4966},
    doi = {10.1121/1.4707424},
    url = {https://doi.org/10.1121/1.4707424},
    eprint = {https://pubs.aip.org/asa/jasa/article-pdf/131/6/4640/15300918/4640_1_online.pdf},
}

@InProceedings{troshani2024weak,
author="Troshani, Ilira
and Gouv{\^e}a, Thiago S.
and Sonntag, Daniel",
editor="Hotho, Andreas
and Rudolph, Sebastian",
title="Leveraging Weakly Supervised and Multiple Instance Learning for Multi-label Classification of Passive Acoustic Monitoring Data",
booktitle="KI 2024: Advances in Artificial Intelligence",
year="2024",
publisher="Springer Nature Switzerland",
address="Cham",
pages="260--272",
isbn="978-3-031-70893-0"
}

@inproceedings{cui2020multisource,
    title = "Multi-Source Attention for Unsupervised Domain Adaptation",
    author = "Cui, Xia  and
      Bollegala, Danushka",
    editor = "Wong, Kam-Fai  and
      Knight, Kevin  and
      Wu, Hua",
    booktitle = "Proceedings of the 1st Conference of the Asia-Pacific Chapter of the Association for Computational Linguistics and the 10th International Joint Conference on Natural Language Processing",
    month = dec,
    year = "2020",
    address = "Suzhou, China",
    publisher = "Association for Computational Linguistics",
    url = "https://aclanthology.org/2020.aacl-main.87/",
    doi = "10.18653/v1/2020.aacl-main.87",
    pages = "873--883",
}

@inproceedings{whisper,
author = {Radford, Alec and Kim, Jong Wook and Xu, Tao and Brockman, Greg and McLeavey, Christine and Sutskever, Ilya},
title = {Robust speech recognition via large-scale weak supervision},
year = {2023},
publisher = {JMLR.org},
booktitle = {Proceedings of the 40th International Conference on Machine Learning},
articleno = {1182},
numpages = {27},
location = {Honolulu, Hawaii, USA},
series = {ICML'23}
}

@inproceedings{wav2vec2,
author = {Baevski, Alexei and Zhou, Henry and Mohamed, Abdelrahman and Auli, Michael},
title = {wav2vec 2.0: a framework for self-supervised learning of speech representations},
year = {2020},
isbn = {9781713829546},
publisher = {Curran Associates Inc.},
address = {Red Hook, NY, USA},
booktitle = {Proceedings of the 34th International Conference on Neural Information Processing Systems},
articleno = {1044},
numpages = {12},
location = {Vancouver, BC, Canada},
series = {NIPS '20}
}

@Article{day2017heterogeneous,
author={Day, Oscar
and Khoshgoftaar, Taghi M.},
title={A survey on heterogeneous transfer learning},
journal={Journal of Big Data},
year={2017},
month={Sep},
day={26},
volume={4},
number={1},
pages={29},
issn={2196-1115},
doi={10.1186/s40537-017-0089-0},
url={https://doi.org/10.1186/s40537-017-0089-0}
}

@article{teixeira2024effective,
author = {Teixeira, Daniella and Roe, Paul and van Rensburg, Berndt J. and Linke, Simon and McDonald, Paul G. and Tucker, David and Fuller, Susan},
title = {Effective ecological monitoring using passive acoustic sensors: Recommendations for conservation practitioners},
journal = {Conservation Science and Practice},
volume = {6},
number = {6},
pages = {e13132},
doi = {https://doi.org/10.1111/csp2.13132},
url = {https://conbio.onlinelibrary.wiley.com/doi/abs/10.1111/csp2.13132},
eprint = {https://conbio.onlinelibrary.wiley.com/doi/pdf/10.1111/csp2.13132},
year = {2024}
}

@article{kurtin2025cuckoo,
author = {Kurtin, Anna M. and Gómez, Erim and Hussey, Nicole and Noson, Anna and O'Reilly, Megan and Rhinehart, Tessa and Skone, Brandi and Wengappuly, Bella and Boyce, Andy J.},
title = {Passive acoustic monitoring paired with machine learning outperforms playback surveys for a rare and cryptic species, the Black-billed Cuckoo (Coccyzus erythropthalmus)},
journal = {Conservation Science and Practice},
volume = {7},
number = {12},
pages = {e70150},
doi = {https://doi.org/10.1111/csp2.70150},
url = {https://conbio.onlinelibrary.wiley.com/doi/abs/10.1111/csp2.70150},
eprint = {https://conbio.onlinelibrary.wiley.com/doi/pdf/10.1111/csp2.70150},
year = {2025}
}

\end{document}